\documentclass[amsmath]{JHEP3}
\usepackage{graphicx}
\usepackage{epsfig}
\usepackage{amsmath}
\title{Quasiparticle Description of the QCD Plasma,\\
Comparison with Lattice Results at Finite $T$ and $\mu$}
\author{
K.~K. Szab\'o and A.~I. T\'oth\\
 \it Institute for Theoretical Physics, E\"otv\"os University, P\'azm\'any
1, H-1117 Budapest, Hungary}

\date{\today}
\maketitle
\abstract{
We compare our $2+1$ flavor, staggered QCD lattice results with 
a quasiparticle picture. 
We determine the pressure, the energy density, the baryon density, 
the speed of sound and the thermal masses as a function of $T$ and $\mu_B$.
For the available thermodynamic quantities the difference 
is a few percent between the results of the two approaches. 
We also give the phase diagram on the $\mu_B$--$T$ plane and estimate
the critical chemical potential at vanishing temperature.
}

\begin{document}
\vspace*{-8.0cm}
\noindent
\hfill 
\vspace*{7.8cm}

\vspace*{0.2cm}

\section{Introduction}
QCD at high temperatures ($T$) and/or quark chemical potentials
($\mu$) plays an important role in particle physics, since
it describes relevant features of nature in extreme
conditions. According to
the standard picture of QCD, at high $T$ 
and/or $\mu$ there is
a change from a state dominated by hadrons
to a state dominated by partons (quarks and gluons). 
This transition
happened in the early Universe (at essentially
vanishing net baryon density \cite{QMconf:2002}) and probably happens in heavy ion
collisions (at moderate but non-vanishing density)
and in neutron stars (at large density, for which
a rich phase structure is conjectured
\cite{Alford:1997zt,Rapp:1997zu,Rajagopal:2000wf}).
There has been several studies in order to determine the thermodynamic properties of the QCD plasma.
The weak coupling expansion of thermodynamic quantities shows a
poor convergence near the transition temperature
\cite{Arnold:ps,Zhai:1995ac,Braaten:1995jr}.
Better convergence is expected from different improved techniques, e.g. 
screened perturbation theory \cite{Karsch:1997gj,Andersen:2000yj}, or 
a systematic rearrangement of the 
perturbative expansion \cite{Blaizot:1999ip}, or 
using an 
effective field theory in 3D
\cite{Kajantie:2000iz}. 
Note that noticeable deviations ($\approx 10\%$) from the ideal gas
limit is expected upto temperatures as high as 1000 times the critical temperature ($T_c$). 
The lattice approach is applicable to investigate
the plasma upto a several times $T_c$ and at zero chemical potential
\cite{Gottlieb:1996ae,Karsch:2000ps,AliKhan:2001ek}. The phenomenology
of these lattice results has been widely analyzed in the literature. 
Successful quasiparticle descriptions 
\cite{Levai:1997yx,Peshier:2001de,Romatschke}
have been introduced to reproduce the properties of the QCD plasma.
One can extrapolate these $\mu=0$ results to non-zero baryon densities 
by using  
the thermodynamical consistency of the model. Unfortunately, until 
recently it was not possible to compare the $\mu \neq 0$ 
predictions of the quasiparticle model with direct 
lattice calculations. The lack of $\mu \neq 0$
lattice results was a consequence of the sign problem. 
The fermionic determinant in the path integral 
becomes complex for non-vanishing $\mu$-s.
This fact spoils any Monte-Carlo based technique 
used in numerical simulations.

Many suggestions were studied in detail to solve 
the sign problem and to give physical answers on the 
lattice at non-vanishing
chemical potentials. Unfortunately, until recently none of them were successful.
Recent interest in this field was initiated
by the overlap improving multi-parameter reweighting 
method \cite{Fodor:2001au}. The phase diagram and the 
critical point were determined in the $2+1$ flavor
QCD on $N_t=4$ lattices with staggered quarks \cite{Fodor:2001pe}. 
Several groups confirmed the results of Refs. 
\cite{Fodor:2001au,Fodor:2001pe} on the phase diagram  
\cite{Allton:2002zi,deForcrand:2002ci,D'Elia:2002pj}.  Furthermore, it
became possible to determine the equation of state (EoS) at 
finite chemical potentials \cite{Fodor:2002au}, too (for 
a recent review on lattice QCD at non-vanishing
chemical potentials see Ref. \cite{Fodor:2002sd}).  
Thus, it would be of interest to see how these new lattice results 
can be described with the quasiparticle approach. This is the
primary goal of the present paper.

The paper is organized as follows. In Section 2 we briefly summarize 
the quasiparticle approach at $\mu =0$ and show how to extend it 
to non-vanishing chemical potentials. Section 3 compares our most
recent lattice results with
the prediction of the quasiparticle model. In Section 4 we summarize. 

\section{Quasiparticle model}
In order to be self-contained we start with 
a brief review of the 
quasiparticle model \footnote{When writing up this paper a different
phenomenological approach was
compared with lattice results at non-vanishing $\mu$ in Ref. \cite{Letessier:2003}.
The quark-gluon plasma liquid 
model was found to be in close agreement with our lattice data.}
suggested in Ref. \cite{Peshier:2001de}. 
Consider a QCD plasma containing gluons ($g$),
$N_l$ number of ``light'' and $N_h$ number of 
``heavy'' quarks ($l$ and $h$, respectively). These basic
constituents have temperature and chemical 
potential dependent effective masses 
\begin{eqnarray}\label{eq:eff_mass}
m_i^2(T,\mu_i)=m_{0i}^2+\Pi^*_i(T,\mu_i), && i = g,l,h, 
\end{eqnarray}
where the $m_{0i}$-s are the rest masses, $\mu_i$-s are the chemical potentials of partons
and the \(\Pi^*_i\)-s are the asymptotic 
values of the hard thermal/density-loop self-energies 
\begin{eqnarray}
\Pi^*_q=2\omega_q (m_0+\omega_q), \quad \quad \omega_q^2=\frac{N_c^2-1}{16N_c}\left(T^2+\frac{\mu_i^2}{\pi^2}\right)G^2, \\ 
\Pi^*_g=\frac{1}{6}\left[\left(N_c+\frac{N_l+N_h}{2}\right)T^2+\frac{3}{2\pi^2} \sum_q \mu_i^2\right]G^2,
\end{eqnarray}
where $G^2$ is the effective gauge coupling (depending on $T$, $\mu$, see below), and 
the summation is performed over the quark flavors ($q$).
We are interested in the non-zero ``light'' quark chemical potential region therefore we set $\mu_{h}=0$
for the ``heavy'' quarks. Note that the baryonic chemical potential is three times the ``light'' quark one. 
In our notations $\mu_{l}=\mu = \mu_B/3$.

The thermodynamic potential the 
pressure $p(T,\mu)$ contains contribution
from the ideal pressure of the quasiparticles ($p_i^{ID}$) and from 
the pressure arising from their interactions ($B(T,\mu)$):
\begin{eqnarray}
p(T,\mu)=\sum_i p^{ID}_i(T,\mu_i(\mu),m_i^2)-B(T,\mu).
\end{eqnarray} 
The ideal gas pressure is the usual Fermi or Bose integral which
takes into account the antiparticles in the quark pressure, too:
\begin{eqnarray}
p^{ID}_{q}(T,\mu)=\frac{d_{q}}{6\pi^2}\int_{m_{q}(T,\mu)}^\infty d\epsilon \frac{(\epsilon ^2-m_{q}^2)^{3/2}}{\exp[(\epsilon-\mu_i)/T]+1}+(\mu_q \to -\mu_q), \\ 
p^{ID}_g(T,\mu)=\frac{d_g}{6\pi^2}\int_{m_g(T,\mu)}^\infty d\epsilon \frac{(\epsilon ^2-m_g^2)^{3/2}}{\exp(\epsilon/T)-1}.
\end{eqnarray}
We do not fix the quark and gluon state multiplicities separately, 
but we demand that their ratio should be 
\(d_g/(d_{l,h})=2\cdot8/(2\cdot3\cdot N_{l,h})\). 
If we impose a stationarity condition \cite{LeeYang:1960} on the pressure then 
further thermodynamic quantities (such as energy 
density ($\epsilon$), quark number density ($n_i$) and entropy density ($s$)) 
will have only quasiparticle and mean field, $B(T,\mu)$, contributions:
\begin{align}
\epsilon(T,\mu)=\sum_i \epsilon^{ID}_i[T,\mu_i(\mu),m_i^2]+B(T,\mu), && 
n_i(T,\mu)=n^{ID}_i[T,\mu_i(\mu),m_i^2],
\end{align}
where $n^{ID}$ is the ideal quark number density.
The functional form of the interaction pressure, 
$B(T,\mu)$, follows from the above condition, too.
The derivatives of $B(T,\mu)$ in $T$ and $\mu$ directions 
are easily accessible quantities 
\begin{eqnarray}
\left.\frac{\partial B}{\partial T}\right|_{\mu}=\sum_i \frac{\partial p^{ID}_i}{\partial m_i^2} \frac{\partial \Pi^*_i}{\partial T}, &&
\left.\frac{\partial B}{\partial \mu}\right|_{T}=\sum_i \frac{\partial p^{ID}_i}{\partial m_i^2} \frac{\partial \Pi^*_i}{\partial \mu},
\end{eqnarray}
so $B(T,\mu)$ can be obtained by an appropriate line integral 
on the $(T,\mu)$ plane.

What remains to be done is to determine the effective 
gauge coupling which appears in
the formula of the self-energies. For vanishing chemical 
potentials it decreases logarithmically
with increasing temperature. A renormalization group inspired parametrization is as follows:
\begin{eqnarray}\label{Gmu=0}
G^2(T,\mu=0)=\frac{48\pi^2}{[33-2(N_l+N_h)]
\log(\frac{T+T_s}{T_c/\lambda})}.
\end{eqnarray}  
Imposing the Maxwell-relation between the 
derivatives of the quark number density and
entropy 
\begin{equation}
\left.\frac{\partial s}{\partial \mu}\right|_{T}=\left.\frac{\partial n_l}{\partial T}\right|_{\mu} \Longrightarrow 
\sum_i\frac{\partial s^{ID}_i}{\partial m_i^2}
\frac{\partial \Pi^*_i}{\partial \mu}=
\frac{\partial n_l^{ID}}{\partial m_l^2}\frac{\partial \Pi^*_l}{\partial T}
\end{equation}
yields a first order, linear partial differential equation for \(G^2(T,\mu)\) 
with straightforwardly calculable \(a_T,a_\mu,b\) 
coefficients 
\begin{equation}\label{de}
a_T(T,\mu,G^2)\cdot\partial _T G^2+a_\mu(T,\mu,G^2) \cdot \partial _\mu G^2=b(T,\mu,G^2).
\end{equation}
This differential equation should be solved with the boundary 
condition at \(\mu=0\) (eq. (\ref{Gmu=0})). Thus,  
the quasiparticle model is unambiguously defined by using 
thermodynamical consistency for non-zero $\mu$-s above the critical line ($T_c(\mu)$).
Below $T_c(\mu)$ the solution of the differential equation, thus the pressure is not unique. 
In this region the system has hadronic degrees of freedom instead of partonic. The 
quasiparticle model constructed from partons looses its validity.

\section{Comparison with lattice results}

As we have already mentioned the sign problem 
of lattice QCD at finite chemical potentials spoils
any Monte-Carlo method based on importance sampling.
The recently proposed overlap ensuring multi-parameter 
reweighting method \cite{Fodor:2001au} enabled us 
to determine the EoS at non-zero temperatures
and chemical potentials \cite{Fodor:2002au}. The 
simulations were carried out on
$N_t=4$ temporal and $N_s=2N_t\dots 3 N_t$ spatial 
extension lattices with two ``light'' ($m_l=0.384 T_c$)
and one ``heavy'' ($m_h\approx T_c$) dynamical quarks.
Note that the mass of the ``heavy'' quark  
corresponds approximately to 
the physical mass of the strange quark, 
whereas the mass of the
``light'' quarks is several times larger 
than the physical values for the up/down quarks.
In the lattice analysis
both the temperature and the baryon chemical potential 
covered the range upto $\mu_B\approx 3 T_c$.

Our lattice calculations were done on lattices with $N_t=4$ temporal
extension. In order to help the continuum interpretation
we multiply the lattice results by 
the dominant $T\rightarrow\infty$ correction factors 
between the $N_t=4$ and the continuum case. We denote the 
$T\rightarrow\infty$ limit as the SB (Stefan-Boltzmann) case.
\begin{eqnarray}
c_p=\frac{p^{SB}(\mu=0,T\to \infty,\rm{continuum})}
        {p^{SB}(\mu=0,T \to \infty,N_t=4)}=0.518, \\
c_\mu=\frac{\Delta p^{SB}(\mu,T \to \infty,\rm{continuum})}
{\Delta p^{SB}(\mu,T \to \infty,N_t=4)}=0.446 + {\cal O}\left(\frac{\mu^2}{T^2}\right).
\end{eqnarray}
Note that the $m$ dependence of these factors are suppressed in the $T\to \infty$ limit.
Here $\Delta p$ indicates the difference between the pressure at $\mu\neq 0$ and the 
pressure at $\mu=0$.
The well-known continuum expressions for the non-interacting quark gluon plasma are 
\begin{align}
p^{SB}(\mu=0,T\to \infty,\rm{continuum})=[16+21(\it{N_l}+\it{N_h})\rm{/2}]\pi^2\it{T}\rm{^4/90}, \\
\Delta p^{SB}(\mu,T \to \infty,\rm{continuum})=\it{N_l}\rm{\mu^2}\it{T}\rm{^2/2+{\cal O}(\mu^4).} 
\end{align}
Including these dominant multiplicative corrections
the results 
might be interpreted as continuum estimates. (Clearly,
the appropriate -- but more CPU-consuming -- way is to
carry out the lattice calculations at $N_t>4$ and perform
an extrapolation to $N_t\rightarrow\infty$.)

The quasiparticle model has four free parameters:
$\lambda$ and $T_s$ to determine the gauge coupling,
$d_g$ the gluon multiplicity and $B(T_c)$ the integration constant in
the interaction pressure. They were used as fit parameters in order to receive 
the least possible difference between the thermodynamic observables measured on the lattice
and the ones predicted by the quasiparticle description.  
In principle it is sufficient to use only the lattice results at vanishing $\mu$ 
in the fitting procedure since afterwards the thermodynamical consistency unambigously
defines the quasiparticle approach; but this could lead to large differences 
between the two types of descriptions 
at non-vanishing $\mu$-s. It turned out to be more sensible to use both the $\mu=0$ and the $\mu\neq 0$
lattice data for fitting in order to gain a better agreement between the two approaches for 
higher $\mu$-s. Therefore we fitted the quasiparticle picture on the $(\epsilon-3p)/T^4$ at $\mu=0$ and
$\Delta p/T^4$  at our highest chemical potential ($\mu_B\approx 490 \rm{~MeV}$) lattice results, simultaneously. 
The chi-square function ($\chi ^2$) was constructed by
using the statistical errors of the lattice data.
We constructed statistical errors for the fit parameters, which are to give the $68.3\%$ confidence level 
corresponding to the bounds of the $\Delta \chi ^2=4.72$ interval \cite{numrec}. 
Clearly, the 
parameter region according to this confidence level has a complicated shape in the
four-dimensional 
parameter space. The quoted errors just indicate the borders of this domain. 
Furthermore, these errors should be taken with a grain of salt, since
the constructed $\chi ^2$ functions contains no information about the 
systematical uncertainties of our $N_t=4$ lattice data. 
We found it to be more reliable to fit on $(\epsilon-3p)/T^4$ lattice results instead of
$p/T^4$ at $\mu=0$, since minimizing the $\chi ^2$ function of the pressure yielded
an interaction measure without a turning-point around $1.3T_c$, which is typical of the lattice results.  
The best parameter values with statistical uncertainties can be found in
Table \ref{tb:1}. 

\renewcommand{\arraystretch}{1.2}
\begin{table}[h!]
\begin{center}
\begin{tabular}{|c|c|}
\hline
$\lambda$ & $10.1^{\> \> +0.9}_{\> \> -2.0}$ \\
\hline
$T_s/T_c$ & $-0.85^{\> \> +0.06}_{\> \> -0.02}$\\
\hline
$d_g$ & $16.4^{\> \> +0.3}_{\> \> -0.2}$\\
\hline
$B^{1/4}(T_c)$ & $174.3^{\> \> +9.2}_{\> \> -5.2}$ MeV\\
\hline
\end{tabular}
\end{center}
\caption{\label{tb:1}
The best fit parameters of the quasiparticle 
model to our lattice results. 
}
\end{table}

Using the best fit parameters we can obtain the temperature and chemical potential dependence of several 
thermodynamic quantities and compare them with lattice data. 
Note that we use $T_c=170 \rm{~MeV}$ as the overall scale in the results.
In formulas and figures $T_c$ is used as the transition temperature
at vanishing chemical potential, whereas $T_c(\mu_B)$ indicates the transition
temperature at non-vanishing chemical potentials. We chose the chemical potential values  
so as to cover the characteristic $\mu$-s occuring in the heavy ion collisions. At our smallest chemical potential value, 
$27$~MeV, and at around $300$~MeV Au+Au colliding experiments were carried out at RHIC \cite{Andronic:2002dec,Csorgo:2002jul}, 
whereas at $253$~MeV Pb+Pb collisions took place at SPS \cite{Broniowski}.       
\begin{figure}
\begin{center}
\includegraphics*[width=7.0cm,bb=20 190 570 710]{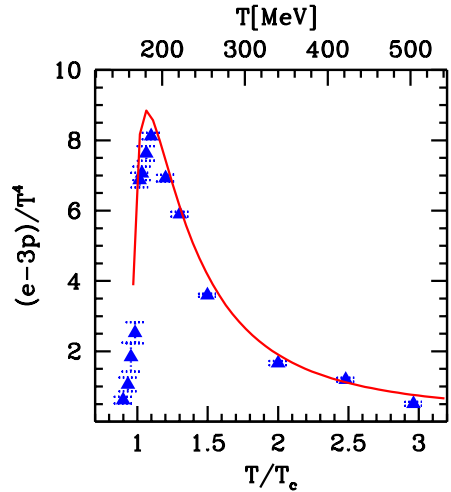}
\end{center}
\caption{\label{fi:eos0_en}
The interaction measure, ($\epsilon-3p$), 
normalized by $T^4$ as 
a function of $T/T_c$ at $\mu=0$. The line corresponds to 
the quasiparticle model, the points are lattice results multiplied 
by $c_p$. The errorbars show the statistical uncertainties. 
}
\end{figure}
\begin{figure}
\begin{center}
\includegraphics*[width=7.0cm,bb=20 190 570 700]{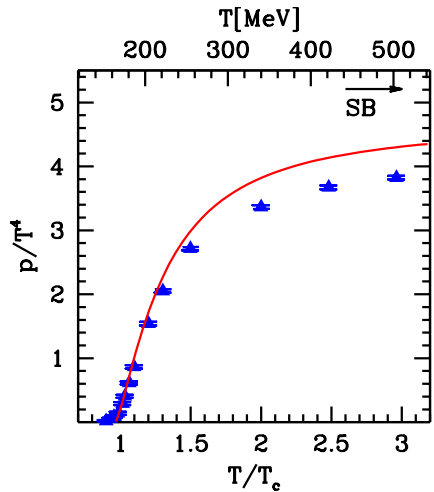}
\end{center}
\caption{\label{fi:eos0_p}
The pressure normalized by $T^4$ as a function 
of $T/T_c$ at $\mu=0$. The line corresponds to the quasiparticle model,
the points are lattice results after a multiplication by $c_p$. 
The arrow indicates the high temperature 
ideal gas pressure of the QCD plasma (SB limit).
}
\end{figure}
\begin{figure}
\begin{center}
\includegraphics*[width=7.0cm,bb=20 170 570 700]{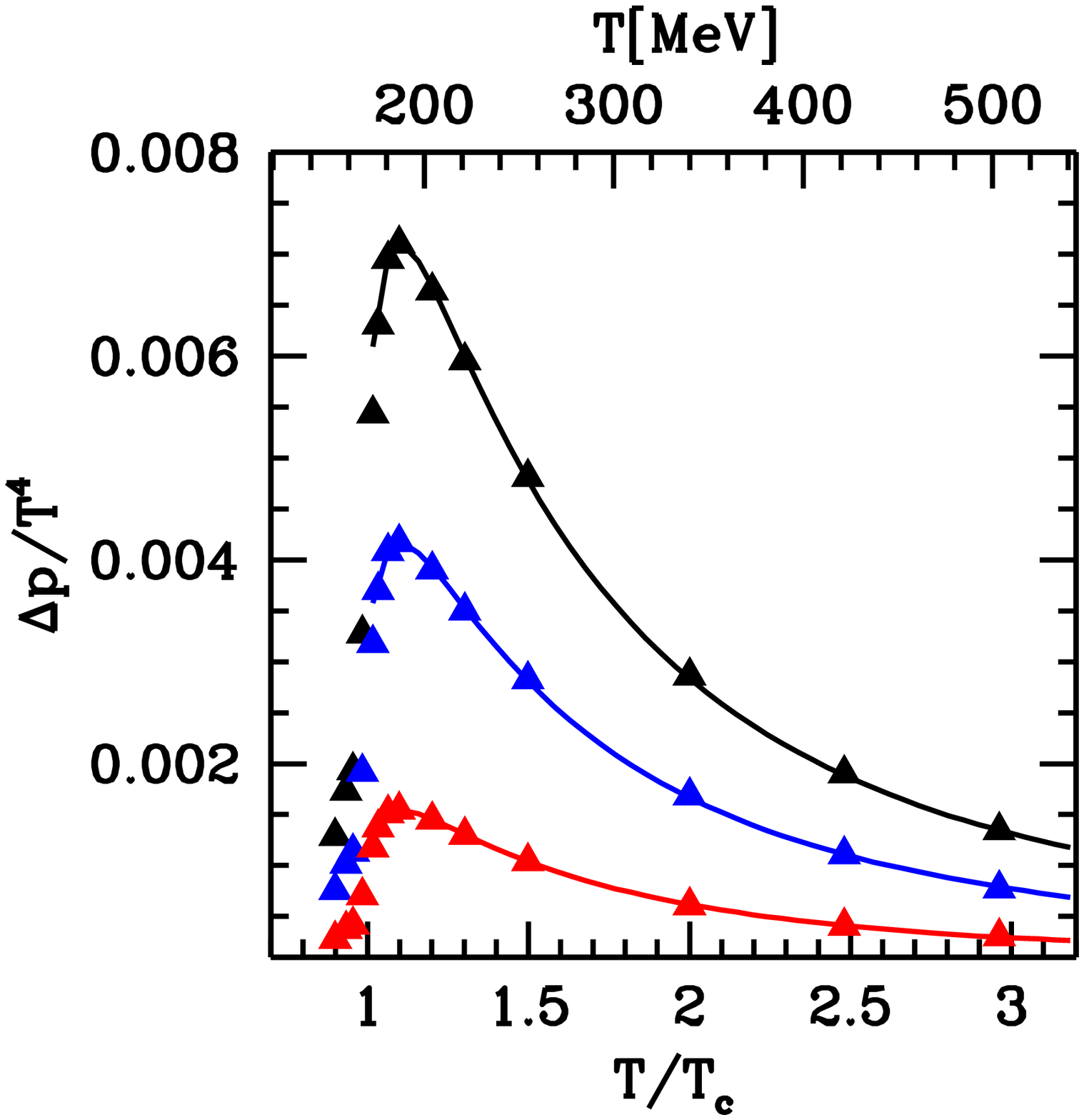}
\includegraphics*[width=7.0cm,bb=20 170 570 700]{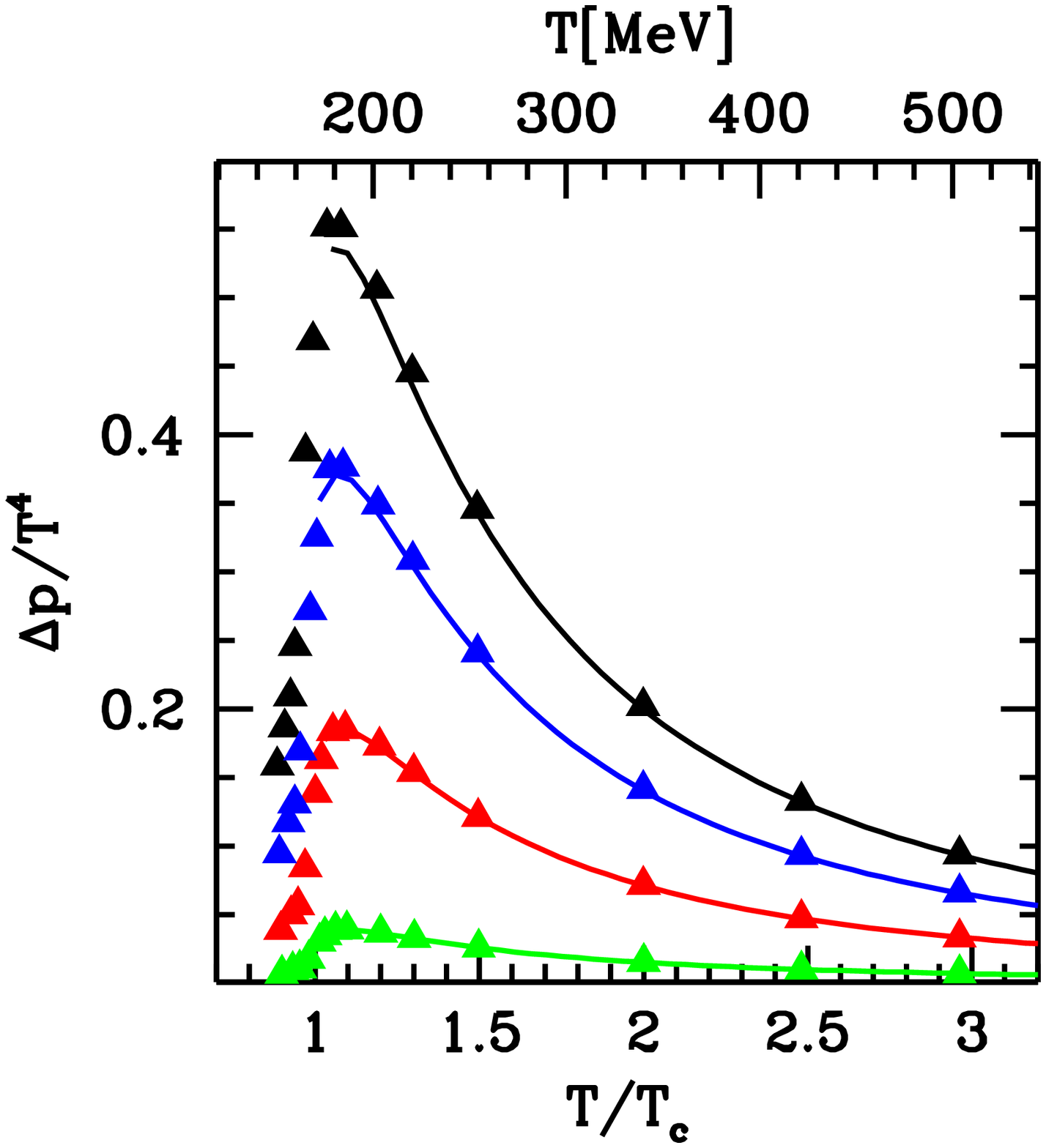}
\end{center}
\caption{\label{fi:dp}
$\Delta p=p(\mu\neq 0,T)-p(\mu=0,T)$ normalized by $T^4$ as a function of $T/T_c$ 
for $\mu_B=27,45$ and $60$~MeV in the left panel and 
for $\mu_B=140,290,410$ and $490$~MeV in the right panel (upper curves correspond to 
larger chemical potentials). The lines represent  
the quasiparticle model, the points are lattice results multiplied 
by $c_\mu$.  The statistical uncertainties
are smaller then the point size. 
}
\end{figure}
\begin{figure}
\begin{center}
\includegraphics*[width=7.0cm,bb=20 190 570 700]{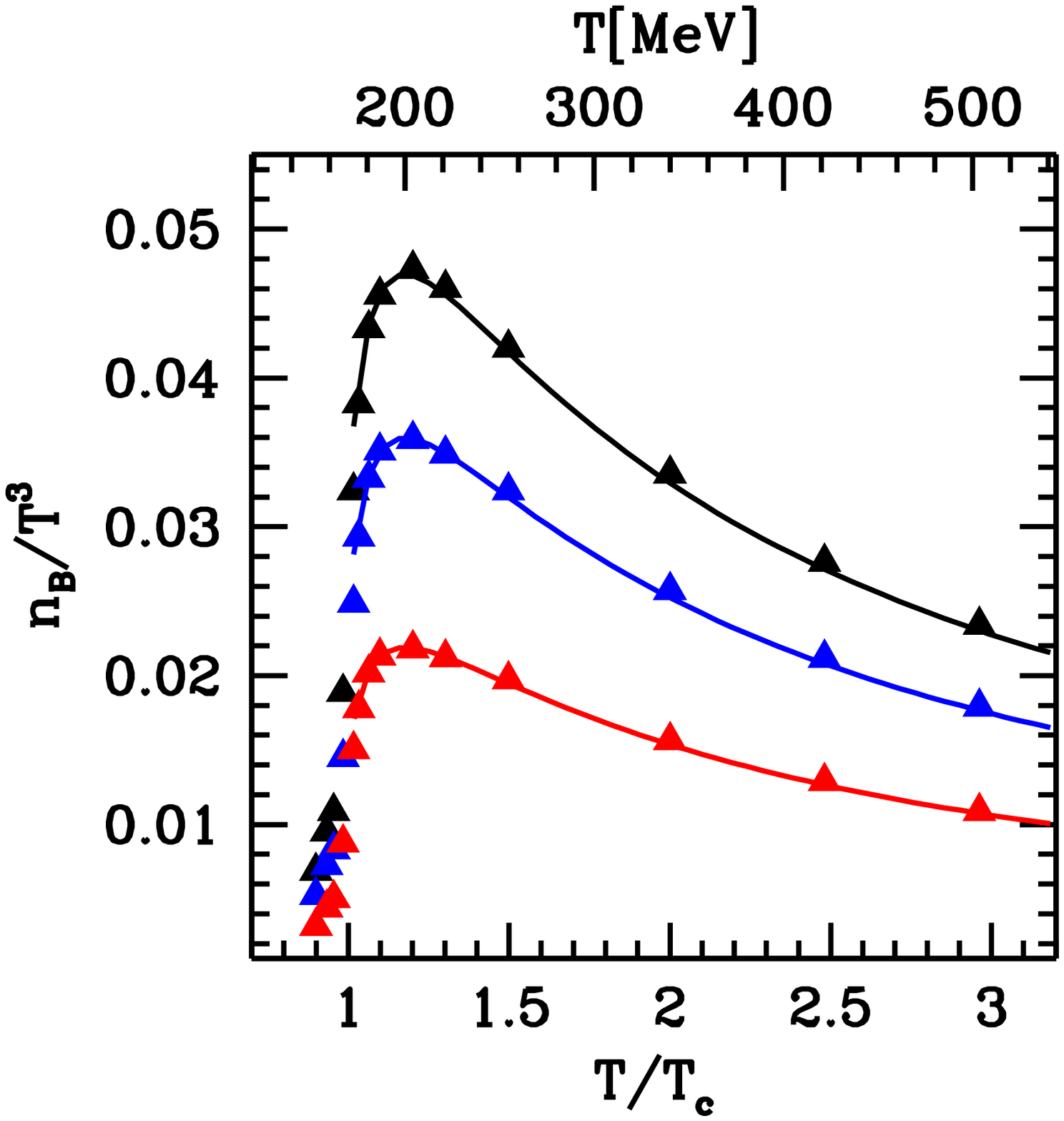}
\includegraphics*[width=7.0cm,bb=20 190 570 700]{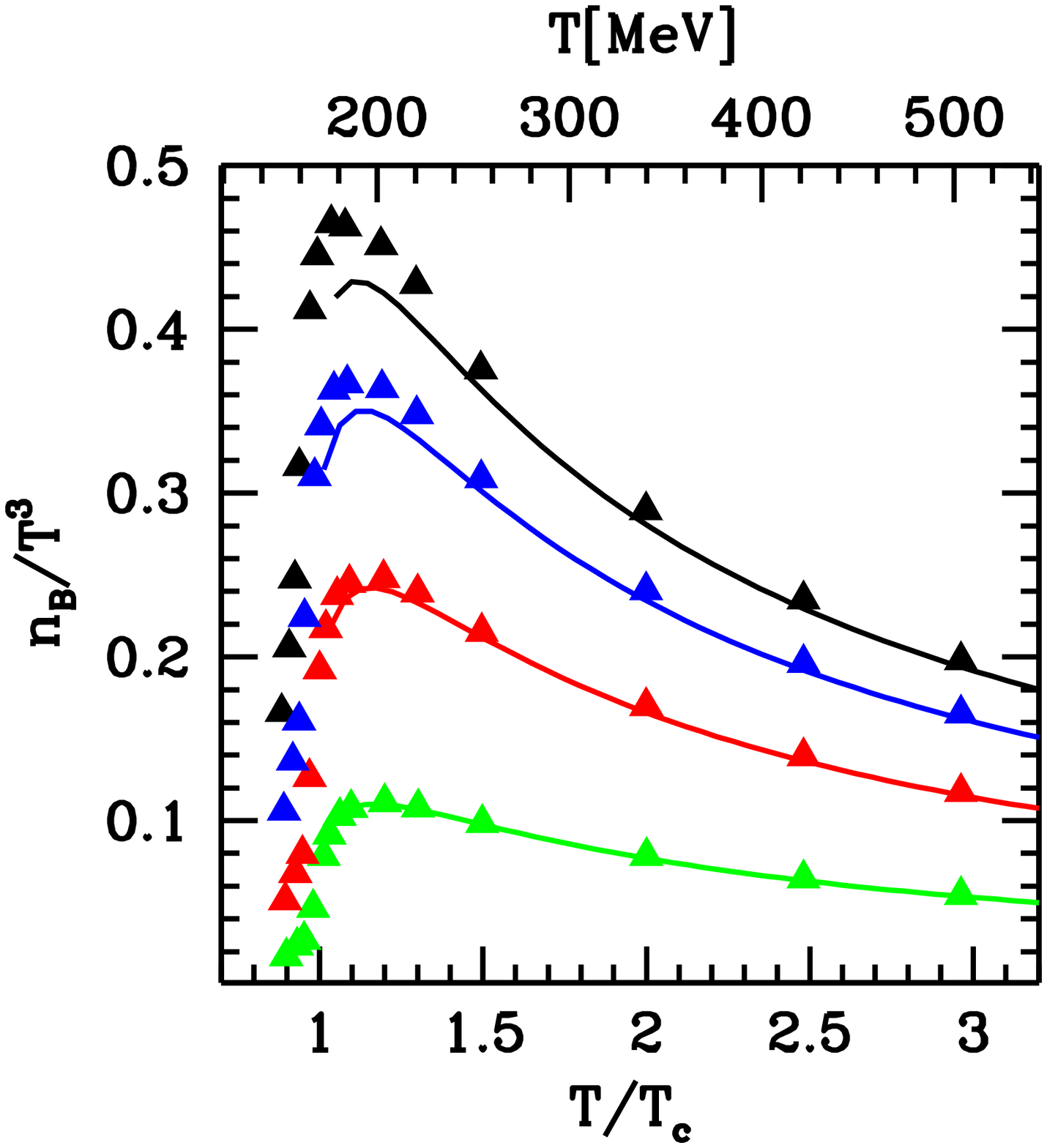}
\end{center}
\caption{\label{fi:n_B}
The baryon number density normalized by $T^3$
as a function of $T/T_c$ for $\mu_B=27,45$ and $60$~MeV in the left panel and 
for $\mu_B=140,290,410$ and $490$~MeV in the right panel
(upper curves correspond to 
larger chemical potentials). The lines represent    
the quasiparticle model, the points are lattice results multiplied
by $c_\mu$.
The statistical uncertainties
are smaller then the point size.
}
\end{figure}
\begin{figure}
\begin{center}
\includegraphics*[width=7.0cm,bb=20 170 570 700]{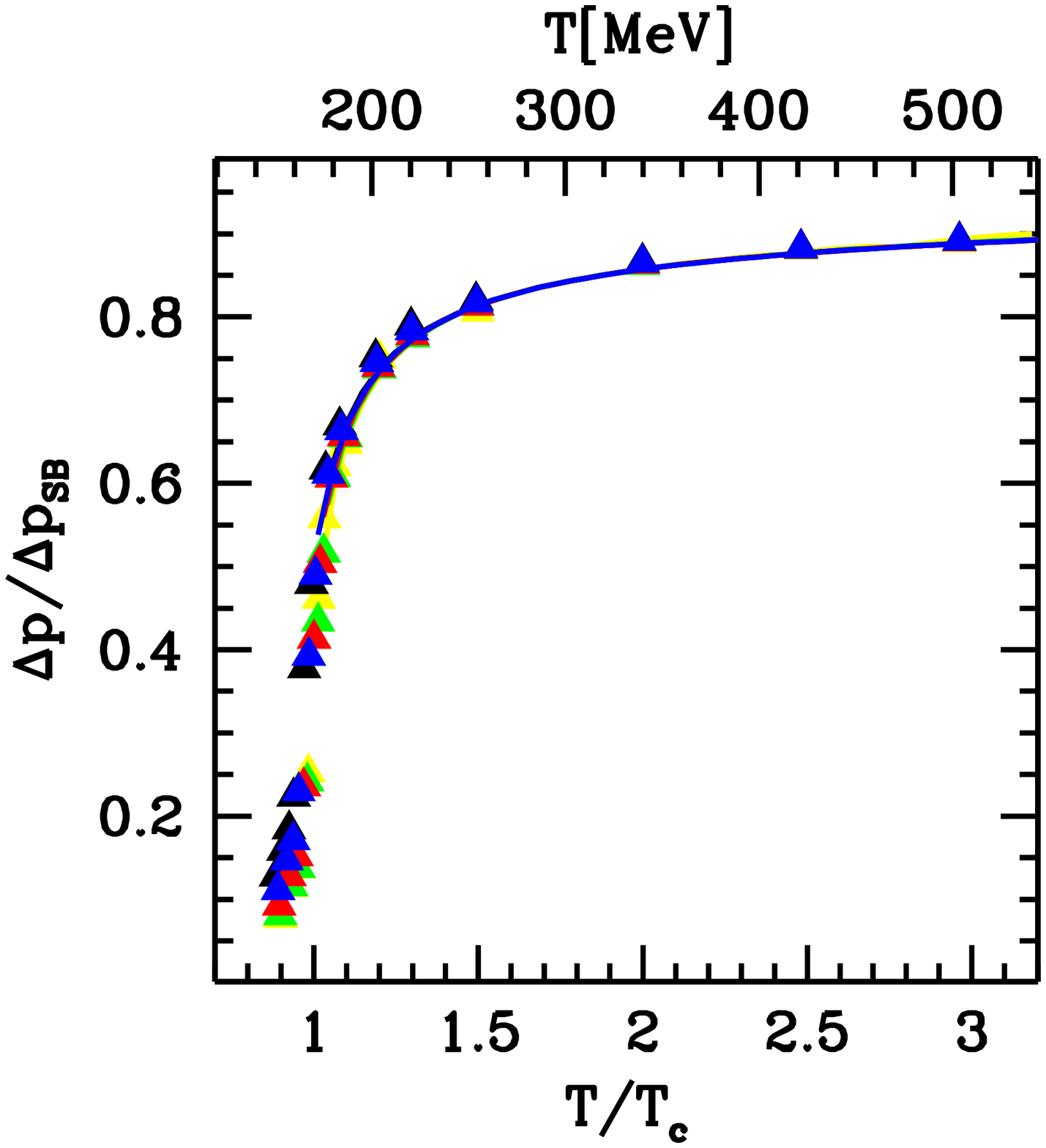}
\end{center}
\caption{\label{fi:dp_SB}
$\Delta p$ of the interacting QCD
plasma normalized by $\Delta p^{SB}$ of the free gas 
(SB) as a function of $T/T_c$ for $\mu_B=60,140,290,410$ and $490$~MeV. 
The lines represent  
the quasiparticle model, the points are lattice results.
The statistical uncertainties
are smaller then the point size.
}
\end{figure}

Figures \ref{fi:eos0_en} and \ref{fi:eos0_p} show the behavior of the ``interaction 
measure'', ($\epsilon-3p$), and the pressure (both normalized by $T^4$) at vanishing
chemical potentials. Both the lattice result and the prediction of the quasiparticle approach
are given. On the one hand there is a good agreement between the two 
techniques for ($\epsilon-3p$), on the other hand for the pressure one finds
a larger difference. It is easy to understand the reason for that. First of
all we used $(\epsilon-3p)$ in our fitting procedure, therefore a better
agreement is expected for this quantity than for $p$, which is a 
predicition of the model. Secondly, the
pressure can be obtained as an integral of the interaction measure
\begin{equation}\label{integral}
\frac{p}{T^4}=\int\frac{dT}{T}\frac{\epsilon-3p}{T^4}.
\end{equation}
Differences in ($\epsilon-3p$) at small temperatures dominates the above
integral. Though the difference
between the lattice and the quasiparticle results for ($\epsilon-3p$) 
is rather small at large $T$  
one observes a 20\% difference around $T_c$.
Therefore we end up with an
$\approx 10\%$ difference for $p$ even in the large $T$ region. 
There might be several
explanations for the differences between the two approaches. 
It can be that the lattice results
in the continuum limit ($N_t\rightarrow\infty$) change the picture
and lead to smaller
discrepancies. Alternatively, it also can be that the  
quasiparticle approach
simplifies the interaction in the quark-gluon plasma. This means
that the predicition of this method can be $\approx 10\%$ off.

We show the temperature dependence of $\Delta p/T^4 $ in Figure \ref{fi:dp} 
and that of $n_B/T^3 $ in Figure \ref{fi:n_B}. 
Both figures indicate a nice agreement between the two types of QCD plasma
descriptions (the largest difference is  $\approx 5 \%$ at 
low temperatures and high chemical potential). 
The temperature dependence of the baryon density is similar 
to that of the pressure since they are connected by the formula
\begin{equation}
T\left.\frac{\partial}{\partial \mu_B}\right|_T\left(\frac{\Delta p}{T^4}\right)=\frac{n_B}{T^3}.
\end{equation}

We observe an interesting scaling behavior of 
$\Delta p$ normalized by $\Delta p^{SB}$
(Figure \ref{fi:dp_SB}).
This quantity ($\Delta p/\Delta p^{SB}$) 
depends only on $T/T_c$ and it is almost independent 
of the chemical potential. This scaling behaviour is less accurate
around $T/T_c$ and it gets more and more precise for 
higher temperatures.  

It is of extreme importance to determine the phase line
(the line, which separates the phases dominated by hadrons and 
partons). One is particularly interested in results
at low temperatures, for which only model estimates are
available. First we use the quasiparticle picture, then an
extrapolation to the lattice data in order to determine the 
phase line. 
\begin{figure}[h!]
\begin{center}
\includegraphics*[width=7.0cm,bb=20 190 570 700]{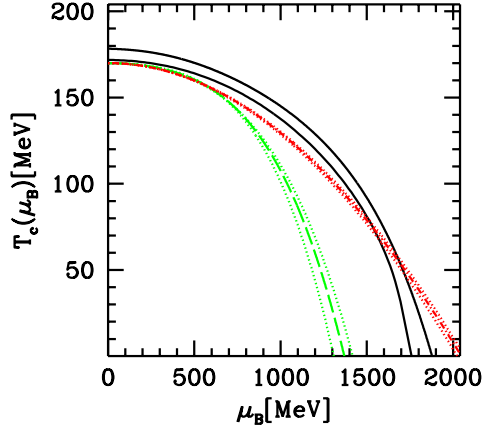}
\end{center}
\caption{\label{fi:tcmu}
The critical lines on the temperature versus chemical potential plane. 
The solid curves are   
quasiparticle critical lines which were obtained with the best
fit parameters. The lower curve corresponds to the pressure equals to zero criteria while 
the pressure equals to $0.14p^{SB}$ along the upper curve. 
The dashed and dashed dotted (lower and upper) 
lines are the extrapolations of the lattice critical line to 
$T=0$ using a 4th and a 2nd order polynom, 
respectively. 
The dotted lines show the influence of the 
statistical uncertainties of the lattice results.
Note that our critical lines loose their validity towards higher chemical potentials,
where the color superconducting phase comes into play.
}
\end{figure}
\begin{figure}[t!]
\begin{center}
\includegraphics*[width=7.0cm,bb=20 190 570 700]{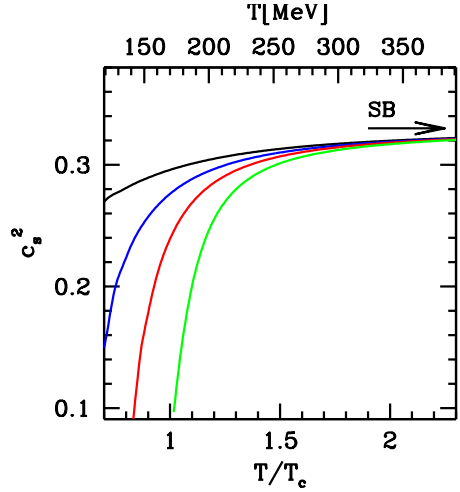}
\end{center}
\caption{\label{fi:cs}
The speed of sound squared in the QCD plasma as a function of $T/T_c(\mu_B)$ at various chemical potentials 
($\mu_B=0,900,1200$ and $1500$~MeV). Upper curves correspond to larger 
values of $\mu_B$.
The arrow indicates the speed of sound squared in the ideal 
gas limit ($(c_s^{SB}) ^2=1/3$).
}
\end{figure}

We defined a quasiparticle critical line based on the following observation (see Figure\ref{fi:eos0_p}):
the pressure around $T_c$ is a lot lower than the pressure at high temperatures, which is in the order of $p^{SB}$.
A simple explanation of this fact is that the number of degrees of freedom in the QGP is much bigger 
than in the hadronic phase, which is a dilute gas of pions.
So we set $p$ to be a few percent of $p^{SB}$ along the phase line (note that the 
result is rather insensitive to the exact
value of this percentage as illustrated in Figure \ref{fi:tcmu}). 
The line defined above is depicted also at rather low temperatures and we 
get $\approx 1800$ MeV for the point of intersection on the $T=0$ axes.
We must interpret this value as the critical chemical potential ($\mu_B^{crit}$) with great care
since for the low temperature, high chemical potential region the 
color-superconducting phase is conjectured \cite{Alford:CSC}, which is obviously not contained in this 
quasiparticle framework.  
\begin{figure}[t!]
\begin{center}
\includegraphics*[width=7.0cm,bb=20 190 570 700]{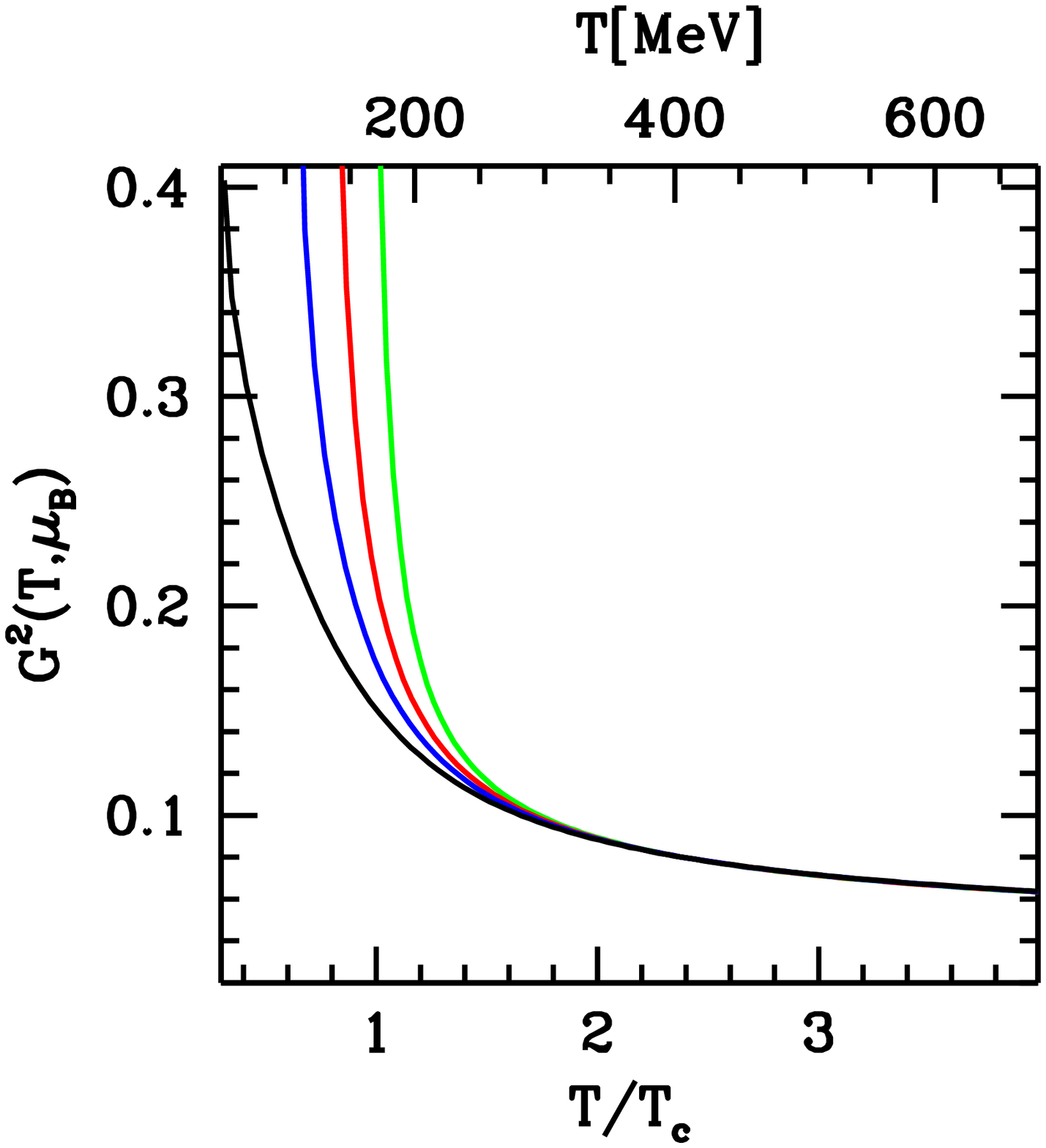}
\end{center}
\caption{\label{fi:g2}
The coupling constant squared as a 
function of $T/T_c$ for various chemical potentials 
($\mu_B=0,900,1200$ and $1500$~MeV). 
Upper curves correspond to smaller 
values of $\mu_B$.  
}
\end{figure}
The quasiparticle transition line 
shows a nice agreement with the directly measured
lattice transition line in the 
$\mu_B<3 T_c$ region (Figure \ref{fi:tcmu}). Outside of this region we do not have direct lattice
results. It is intriguing to see  
the behavior of the critical lines when $T \rightarrow 0$. 
We extended our lattice transition line to zero temperature 
by using two fits namely a second and a fourth order  
polynom. Due to $\mu$ versus -$\mu$ symmetry we keep only even order 
terms in the fitting procedure. The bands in Figure \ref{fi:tcmu} 
indicate the statistical uncertainties of the fits. The two curves
agree nicely in the directly measured $\mu_B<3 T_c$ region. They
deviate for smaller temperatures and larger chemical potentials.
The quadratic polynomial fit predicts  
at vanishing temperature $\mu_B^{\rm crit} \approx 2000$ MeV
(which exceeds expectations). The quartic approximation
gives for the same quantity $\mu_B^{\rm crit} \approx 1400$ MeV. 
The difference between the 
predicitions of the two kinds of polynoms 
might be interpreted as a rough estimate for the systematic uncertainty 
of the extrapolation to low temperatures. 

The dynamical properties of the plasma phase is 
primarily determined by the speed of sound
\begin{equation}
c_s^2(T,\mu_B)=\frac{dp}{d\epsilon}.
\end{equation}
As it can be seen in Figure \ref{fi:cs} increasing 
chemical potential yields higher speed of sound.

One can also look for the predictions of the quasiparticle model 
for quantities which are straightforward predicitions of the
quasiparticle approach however, it is difficult to obtain them
directly on the lattice.
Such important quantity is the gauge coupling of the quasiparticle picture,
which is shown in Figure \ref{fi:g2}. Note that the static potential
at finite chemical potentials can be in principle measured on the lattice.
Using the lattice potential one can easily define the gauge coupling.
Clearly, the parametrization of the gauge coupling in the 
quasiparticle picture is not equivalent with the above mentioned coupling.

\begin{figure}[t!]
\begin{center}
\includegraphics*[width=7.0cm,bb=20 190 570 700]{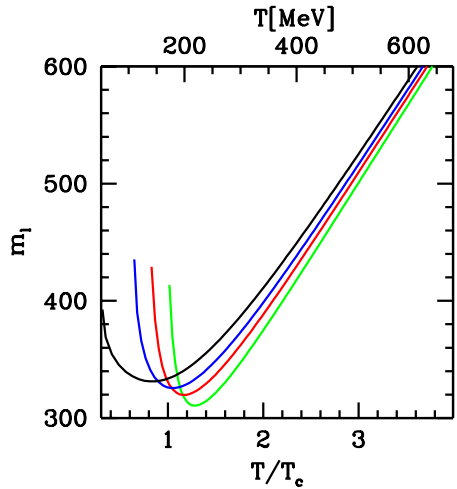}
\end{center}
\caption{\label{fi:mpertq}
The effective mass of the light quarks 
as a function of $T/T_c$ for various chemical potentials
($\mu_B=0,900,1200$ and $1500$~MeV, right to left).
}
\end{figure}

\begin{figure}[t!]
\begin{center}
\includegraphics*[width=7.0cm,bb=20 190 570 700]{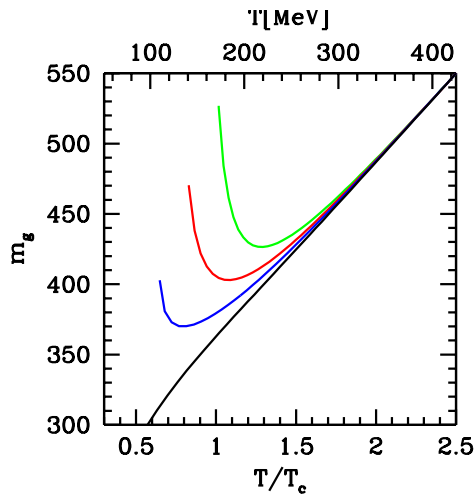}
\end{center}
\caption{\label{fi:mpertg}
The effective mass of the gluon 
as a function of $T/T_c$ for various chemical potentials
($\mu_B=0,900,1200$ and $1500$~MeV).
Upper curves correspond to smaller       
values of $\mu_B$. 
}
\end{figure}

In order to understand the typical degrees of freedom in the high
temperature QCD phase at non-vanishing chemical potentials
it is instructive to study
the effective masses of the quasiparticles.
Figure \ref{fi:mpertq} shows the temperature dependence of the light
quark masses for different chemical potentials, whereas Figure \ref{fi:mpertg}
presents the results on the gluon masses. 
According to our observations these quantities
are almost independent of the chemical potential 
in the $\mu_B<3 T_c$ region. Due to the 
stationarity condition $\partial p/\partial m_{\rm i}=0$
small changes in the mass do not change the pressure.
It means that the chemical potential dependence of the thermodynamic 
quantities are primarily coming from the direct $\mu_B$ 
dependence of the Fermi integrals. 

\section{Conclusion}
In this paper we studied the quasiparticle approach
to describe the equation of state of the hot QCD plasma.
We have fitted the free parameters of the model by using
our $2+1$ flavor, dynamical staggered QCD lattice results in the region 
$T, \mu_B<3 T_c$. After calculating 
the pressure, interaction measure and density 
we found 
a good agreement between the quasiparticle
predictions and our lattice data. 
The model successfully justifies the scaling behavior of $\Delta p/\Delta p^{SB}$
observed in the lattice calculations.
We gave some confidence intervals for the 
fit parameters in spite of the lack of continuum extrapolations. 
Using the best fit parameters the quasiparticle critical line and the speed of sound 
were given for higher values of $\mu_B$. The zero temperature limit of the quasiparticle and
lattice critical lines (critical $\mu_B$ value) cover the $\mu_B^{\rm crit} \approx 1400 \dots 2000$ MeV region. 

\vspace{0.5cm}  
\noindent
{\bf Acknowledgements:\\} 
We highly appreciate Z.~Fodor's continuous help. 
Useful suggestions on the manusscript from T.~Cs\" org\H o, 
B.~K\"ampfer, S.~D.~Katz and P.~L\'evai 
are also acknowledged.
This work was partially supported by Hungarian Scientific
grants, OTKA-T34980/\-T29803/\-T37615/\-M37071/\-OMFB1548/\-OMMU-708.

\end{document}